\documentclass[a4paper,fleqn,usenatbib,useAMS]{mnras}

\usepackage{graphicx}	
\usepackage{amsmath}	
\usepackage{amssymb}	
\usepackage{multicol}        
\usepackage{bm}		
\usepackage{pdflscape}	

\usepackage[T1]{fontenc}
\usepackage{ae,aecompl}

\usepackage{newtxtext,newtxmath}

\title[New Detection Method for Ultra-low Frequency GWs]{Sensitivity of new detection method for ultra-low frequency gravitational waves with pulsar spin-down rate statistics}

\author[N. Yonemaru, H. Kumamoto, K. Takahashi \& S. Kuroyanagi]{Naoyuki Yonemaru$^1$\thanks{e-mail: \href{178d9005@st.kumamoto-u.ac.jp}{178d9005@st.kumamoto-u.ac.jp}}, Hiroki Kumamoto$^1$, Keitaro Takahashi$^1$ \& Sachiko Kuroyanagi$^2$
\\
$^{1}$Kumamoto University, Graduate School of Science and Technology, Japan \\
$^{2}$Nagoya University, Graduate School of Science, Japan}

\date{Last updated 2015 May 22; in original form 2013 September 5}

\pubyear{2017}

\begin{document}
\label{firstpage}
\pagerange{\pageref{firstpage}--\pageref{lastpage}}
\maketitle

\begin{abstract}
A new detection method for ultra-low frequency gravitational waves (GWs) with a frequency much lower than the observational range of pulsar timing arrays (PTAs) was suggested in \citet{Yonemaru}. In the PTA analysis, ultra-low frequency GWs ($\lesssim 10^{-10}$ Hz) which evolve just linearly during the observation time span are absorbed by the pulsar spin-down rates since both have the same effect on the pulse arrival time. Therefore, such GWs cannot be detected by the conventional method of PTAs. However, the bias on the observed spin-down rates depends on relative direction of a pulsar and GW source and shows a quadrupole pattern in the sky. Thus, if we divide the pulsars according to the position in the sky and see the difference in the statistics of the spin-down rates, ultra-low frequency GWs from a single source can be detected. In this paper, we evaluate the potential of this method by Monte-Carlo simulations and estimate the sensitivity, considering only the "Earth term" while the "pulsar term" acts like random noise for GW frequencies $10^{-13}-10^{-10}$ Hz. We find that with 3,000 milli-second pulsars, which are expected to be discovered by a future survey with the Square Kilometre Array, GWs with the derivative of amplitude of about $3 \times 10^{-19}~\rm{s}^{-1}$ can in principle be detected. Implications for possible supermassive binary black holes in Sgr${}^*$ and M87 are also given.
\end{abstract}

\begin{keywords}
gravitational waves -- (stars:) pulsars: general --methods: statistical
\end{keywords}




\section{Introduction}

Low frequency gravitational waves (GWs) at $10^{-9} - 10^{-6}$ Hz can be detected with pulsar timing arrays (PTAs) \citep{Foster}. This method utilizes the fact that GWs affect the arrival time of pulses and the signal can be extracted from the timing residuals, which is the deviation of the pulse arrival times from the expectation without GWs. The detectable frequency-range is determined by the observational time span and cadence. Currently, three PTAs are in operation; the Parkes PTA in Australia \citep{Manchester}, the European PTA \citep{Kramer13} and NANOGrav in the United States \citep{Jenet}. Further, in the 2020s, the Square Kilometre Array (SKA) will start running \citep{Kramer15} and 1,400 and 3,000 millisecond pulsars (MSPs) are predicted to be discovered by the SKA1 and SKA2 surveys \citep{Keane}, respectively.

In this frequency range, GW sources are considered to be mainly inspiraling supermassive black hole (SMBH) binaries, cosmic strings and their incoherent superposition, the gravitational wave background (GWB). In the case of SMBH binaries, the frequency range of PTAs corresponds to the sub-pc orbital radii (e.g., $6.31\times 10^{-3}$ pc for an equal-mass of $10^8M_\odot$ binary at $10^{-8}$ Hz), so that only the binaries in the late stage of the evolution can be probed with PTAs. In the early stage of the evolution, the orbit of a SMBH binary shrinks as the angular momentum is extracted by scattering of stars and the friction of surrounding gas. However, when the orbital radius becomes a few pc, the transfer of angular momentum by stars and gas is no longer effective so that it exceeds the Hubble time for two SMBHs to merge by only GW emission \citep{Lodato,Milosavljevi´c}. This is called "the final parsec problem" and, therefore, it is important to observe pc-scale binaries in the earlier stage. M87 is one of candidates for host galaxies which have SMBH binaries with large orbits \citep{Batcheldor}. However, GWs from the binary in M87 have too low frequency to be detected with PTAs.

A method to detect GWBs at ultra-low frequency with millisecond pulsar binaries was suggested in the previous works \citep{Bertotti,Kopeikin1997}. In this method, the difference of the time derivatives of {\it{the orbital periods}} between the expectation and observation can be obtained as the GW effect, and the difference is described by the energy density of the stochastic GWB $\Omega_{\rm{GW}}$. An upper limit for the GWB is given as $\Omega_{\rm GW}h^2 \leq 8.5\times 10^{-4}$ at frequencies $10^{-11}~-~7.1\times 10^{-9}$ Hz, where $h$ is the Hubble constant in units of $100~\rm{km~s^{-1}~Mpc^{-1}}$ \citep{Potapov}.

In our previous work \citep{Yonemaru}, we proposed a new detection method for a single GW with a ultra-low frequency ($\lesssim 10^{-9}~{\rm Hz}$). In a observational time span of order 10 years, the amplitude of ultra-low frequency GWs does not appear like a wave but evolve linearly as a function of time. Such GWs are absorbed by the pulsar spin-down rate in the pulsar parameter fitting since both have the same time dependences on the arrival time of pulses, so that such GWs are undetectable by the conventional PTA. However, the bias on the spin-down rate induced by ultra-low frequency GWs depends on the relative direction of the pulsar and GW source in the sky and, fixing the direction of the GW source, the bias shows a quadrupole pattern in the sky. The basic idea of the method is to divide the pulsars according to the position in the sky and see the difference in the statistics of the spin-down rates. Because the quadrupole pattern in the sky is critical for the method, it cannot be applied to the detection of stochastic GW background directly.

In this paper, we estimate the expected sensitivity of our method quantitatively by simulating the spin-down rate statistics of millisecond pulsars considering future pulsar surveys by the SKA. Assuming the position of a GW source and the polarization angle of the GWs, pulsars are divided into two groups according to their position in the sky. If the GW is strong enough, the difference in the spin-down rate distribution of the two groups exceeds the statistical fluctuation. This is the signal of a ultra-low frequency GW.

This paper is organized as follows. In section 2, we describe the detection principle of a new method for ultra-low frequency GWs. In section 3, we employ simulations to assess the sensitivity of this method and represent its results. In section 4, we give a discussion and summary.

\section{Detection principle}\label{principle}

The timing residual due to GWs, the difference between the actual arrival time of pulses from a pulsar and the expectation without GWs, is given by \citep{Detweiler}
\begin{equation}
r_{\rm{GW}}(t) = \sum_{A = +,\times}F^A(\hat{\Omega},\hat{p})\int^t\Delta h_A(t',\hat{\Omega}, \theta)dt'
\label{residual}
\end{equation}
where $\hat{p}$ and $\hat{\Omega}$ are the direction of the pulsar and GW propagation, respectively, and $\theta$ is the GW polarization angle. Here, $F^A(\hat{\Omega},\hat{p})$ is the antenna beam pattern which is given by \citep{Anholm}
\begin{equation}
F^A(\hat{\Omega},\hat{p})
= \frac{1}{2} \frac{\hat{p}^i\hat{p}^j}{1+\hat{\Omega}\cdot\hat{p}} e^A_{ij}(\hat{\Omega}),
\label{antenna}
\end{equation}
where $e^A_{ij} (A=+,\times)$ is the GW polarization tensor given by
\begin{eqnarray}
e^+_{ij}(\hat{\Omega}) &=& \hat{m}_i\hat{m}_j - \hat{n}_i\hat{n}_j \\
e^\times_{ij}(\hat{\Omega}) &=& \hat{m}_i\hat{n}_j + \hat{n}_i\hat{m}_j,
\end{eqnarray}
where $\hat{m}$ and $\hat{n}$ are the polarization basis vectors. Here $\Delta h_A(t,\hat{\Omega})$ is the difference of the metric perturbation between the earth and pulsar and given by
\begin{equation}
\Delta h_A(t,\hat{\Omega},\theta) = h_A(t,\hat{\Omega},\theta) - h_A(t_p,\hat{\Omega},\theta)\label{dif metric}
\end{equation}
where $t_p = t - \tau$ with $\tau = L/c(1 + \hat{\Omega}\cdot\hat{p})$ is the pulse propagation time from the pulsar to the Earth, and $L$ is the distance to the pulsar. In the following, we will discuss GWs whose period is much longer than the observational time span ($\sim$ 10 years) and, in this case, the GW amplitude changes linearly with time. Further, we neglect the second term (" pulsar term") assuming it has a random phase relative to the first term ("Earth term") for different pulsars and behaves as a random noise with zero mean so that the effect of pulsar term reduces for a large number of pulsars. As we discuss further in section \ref{discussion}, this treatment is valid for GW frequency of $\gtrsim 10^{-13}$ Hz. Then, we can write Eq. (\ref{dif metric}) as
\begin{equation}
\Delta h_A(t,\hat{\Omega},\theta) = \dot{h}_A(\hat{\Omega},\theta)t.\label{hdot}
\end{equation}
Substituting Eq. (\ref{residual}) into Eq. (\ref{hdot}), the timing residual induced by linearly-changing GWs becomes
\begin{equation}
r_{\rm{GW}}(t)
= \frac{1}{2} \sum_{A = +,\times} F^A(\hat{\Omega},\hat{p}) \dot{h}_A(\hat{\Omega},\theta) t^2.
\label{linear res}
\end{equation}
However, such a timing residual is absorbed in the linear-fitting for the timing model of the evolution of the pulse phase in the form of a polynomial of time. Thus, even if such GWs exist, they just contribute to the correction of the quadratic part of the series. Consequently, $\dot{p}/p$ after the polynomial-fitting becomes
\begin{equation}
\frac{\dot{p}}{p} = \frac{\dot{p}_0}{p} + \alpha(\hat{\Omega},\hat{p},\theta),
\end{equation}
where $p$ is the pulse period, $\dot{p}$ and $\dot{p}_0$ are the observed and intrinsic spin-down rates respectively, and $\alpha(\hat{\Omega}, \hat{p}, \theta)$ is a bias factor due to ultra-low frequency GWs which is given by
\begin{equation}
\alpha(\hat{\Omega},\hat{p}, \theta)
= \sum_{A=+,\times} F^A(\hat{\Omega},\hat{p}) \dot{h}_A(\hat{\Omega},\theta).
\label{bias}
\end{equation}
In principle, we cannot separate the GW effect from the intrinsic $\dot{p}_0/p$ for a single pulsar. Therefore, such low-frequency GWs cannot be detected by the conventional PTA method.

\begin{figure}
\centering
\includegraphics[width=60mm, angle=-90]{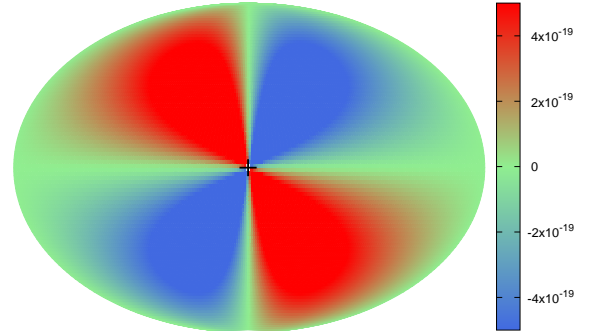}
\caption{Bias factor $\alpha (\hat{\Omega}, \hat{p})$ in the sky for $\dot{h}_+ = 0~\rm{s}^{-1}$ and $\dot{h}_\times = 10^{-18}~\rm{s}^{-1}$. The symbol "+" in the figure represents the source position.}
\label{fig:contour}
\end{figure}

\begin{figure}
\centering
\includegraphics[width = 60 mm,angle=-90]{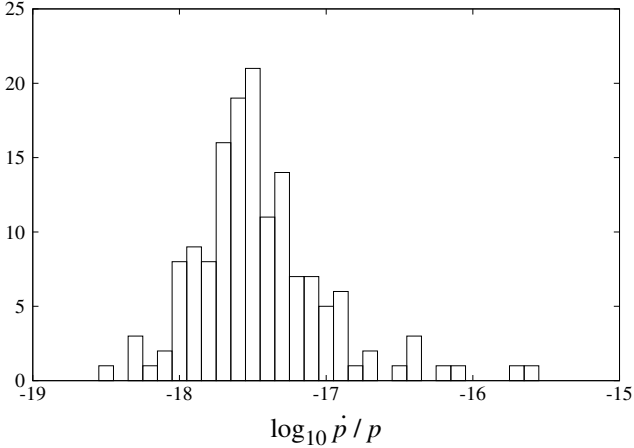}
\caption{Histogram of $\log_{10}\dot{p}/p~[{\rm sec}^{-1}]$ of 149 observed pulsars with $p < 30$ ms. Here pulsars in globular clusters and two ones with negative and extremely large spin-down rates ($\dot{p}/p = -10^{-21.2},~10^{-11.5}~[{\rm sec}^{-1}]$) are not included.}
\label{MSP}
\end{figure}

However, it should be noted that the value of the bias factor, $\alpha(\hat{\Omega},\hat{p},\theta)$, depends on the source position, the GW polarization angle and the pulsar position, and can be either positive or negative. In fact, the distribution of $\alpha(\hat{\Omega},\hat{p},\theta)$ is a quadrupole pattern, as seen in Fig. \ref{fig:contour}. If we divide pulsar samples into two groups according to the sign of the bias factor at the pulsar position, GWs will induce a systematic difference in the spin-down rate distribution of the two groups. Thus, even though the GW signal cannot be extracted from individual pulsars, it will be possible if we utilize the statistics of spin-down rates. However, we note that this method is possible only for a single GW source. In case of multiple sources with comparable $\dot{h}$ values or GWB, the pattern of $\alpha(\hat{\Omega},\hat{p},\theta)$ in the sky is much more complicated and the detection of such GWs would be very hard. Hereafter, we consider only a case of a single source.

Fig. \ref{MSP} shows the histogram of $\log_{10}\dot{p}/p~[{\rm sec}^{-1}]$ of 148 observed MSPs with $p < 30$ ms. Here, MSPs in globular clusters are excluded since they would have been biased significantly by the gravitational potential and complicated dynamics inside the cluster. Further, one MSP with a negative spin-down rate ($\dot{p}/p = -10^{-21.2}~[{\rm sec}^{-1}]$) is not included. Also, one MSP with an extremely large spin-down rate ($\dot{p}/p = 10^{-11.5}~[{\rm sec}^{-1}]$) is excluded as an outlier. We can see that the value of $\log_{10}\dot{p}/p~[{\rm sec}^{-1}]$ ranges from $-18.5$ to $-16$ for "normal" MSPs. Here, we note that the observed spin-down rates are biased by various factors other than GWs: the Shklovskii effect \citep{Shklovskii}, acceleration along the line of sight by gravity inside a globular cluster \citep{Phinney} and the Galactic differential rotation \citep{Damour,Rong}, acceleration toward the Galactic disk \citep{Nice} and low-frequency components of "pulsar timing noise" (or "red noise") often approximated as a power law of the form 
$S(f) \propto f^{-\lambda}$, where $\lambda\ge 1$ is the noise index \citep{Kopeikin1999,Kopeikin2004}. Although the biases from the Galactic differential rotation and acceleration toward the disk could have spatial correlations in the sky, these effects are less significant ($\Delta (\dot{p}/p) \lesssim 10^{-19}$ for MSPs at $\le 10$ kpc) \citep{Nice} and will be removed if the distance to MSPs is measured precisely in the future. We also note that red noise could have a spatial correlation in the case the noise originates in a spatially correlated process such as errors in solar-systems ephemerides \citep{Champion}, or possibly low-frequency components of a GWB \citep{Shannon}. The effects of these biases will be studied elsewhere in the future.

\begin{figure}
\centering
\includegraphics[width=60mm, angle=-90]{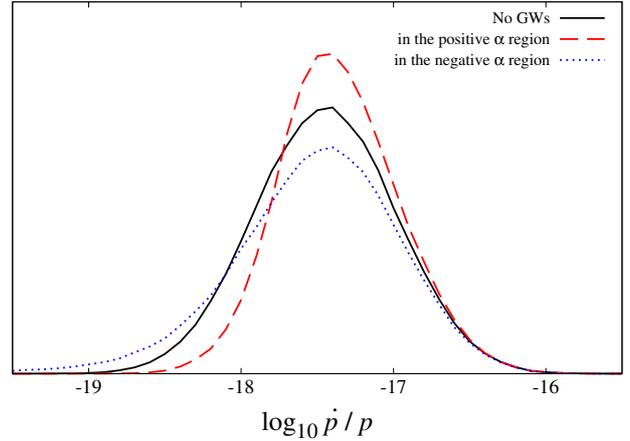}
\caption{Schematic view of the expected systematic difference in the distribution of spin-down rates ($\dot{p}/p$) of two groups in the presence of ultra-low frequency GWs with $\dot{h} = 10^{-18}~\rm{s}^{-1}$. The black line is the assumed intrinsic distribution, which is Gaussian with the mean value and variance which are the same values as the observed ones (Fig. \ref{MSP}). The dashed (red) and dotted (blue) lines show the expected observable distributions in the positive and negative $\alpha(\hat{\Omega},\hat{p})$ regions (the red and blue regions in Fig. \ref{fig:contour}), respectively.}
\label{fig:spin_down}
\end{figure}

Fig. \ref{fig:spin_down} shows a schematic view of the expected systematic difference in the distribution of spin-down rates of two groups mentioned above. In this figure, GWs with $\dot{h} = 10^{-18}~\rm{s}^{-1}$ is assumed and thus pulsars with an intrinsic spin-down rate of this order or smaller ($\dot{p}_0/p \lesssim 10^{-18}~\rm{s}^{-1}$) are significantly affected while those with a much larger value of $\dot{p}_0/p$ are not affected. Overall, the presence of GWs are reflected in the extension of the left-hand-side tail of the observed $\dot{p}/p$ distribution: a short (long) tail for the positive (negative) $\alpha(\hat{\Omega},\hat{p},\theta)$ group. In this paper, we characterize this feature with the skewness of the $\log_{10}\dot{p}/p$ distribution and consider the difference in the skewnesses of the two regions as a statistical measure which reflects the value of $\dot{h}$. 

\section{Simulation results}

In this section, we show results of a series of simulations to evaluate the sensitivity for ultra-low frequency GWs from a single source. The precision of the determination of the source position and GW polarization angle is also investigated.

First, in our simulations, a number of MSPs are located randomly in the sky with an isotropic probability distribution. However, it should be noted that the MSP distribution will not be isotropic but concentrated on the Galactic plane, because the overwhelming majority of the observed MSPs are expected to be Galactic. This means that the sensitivity of the method will be depended on the GW source position. 
Then the spin-down rate $\log_{10}\dot{p}/p$ is given to each of the MSPs according to a Gaussian probability distribution function with the mean and variance of -17.5 and 0.21, respectively, which are taken from the known MSP samples (Fig. \ref{MSP}). Given the direction and polarization angle of assumed GWs, the observed spin-down rates are biased by a factor of $\alpha(\hat{\Omega},\hat{p},\theta)$ depending on the MSP position. After application of the bias factor, we examine the biased spin-down rates and remove any MSPs with negative $\dot{p}$ from the rest of the analysis. In each realization, we create the mock MSP catalogs according to the method described above, and then we perform GW search based on the idea described in the previous section as follows.
 
Assuming the position and polarization angle of GWs, the sky is divided into two regions according to the sign of $\alpha(\hat{\Omega},\hat{p},\theta)$. Then a histogram of observed values of $\log_{10}\dot{p}/p$ is obtained for each region and the difference of skewness between the two histograms is computed, which is defined as
\begin{equation}
\Delta S = S_{\alpha +} - S_{\alpha -},
\end{equation}
where $S_{\alpha +}$ and $S_{\alpha -}$ represent the skewness of the $\log_{10}\dot{p}/p$ distributions in the positive and negative $\alpha(\hat{\Omega},\hat{p},\theta)$ regions, respectively. The skewness is given by
\begin{equation}
S_{\alpha +(-)} = \frac{1}{\sigma^3_{+(-)}N_{+(-)}}\sum^{N_{+(-)}}_i \left( \log_{10}\left(\frac{\dot{p}}{p}\right)_i - \mu_{+(-)} \right)^3,
\end{equation} 
where $i = 1,\cdots,N_{+(-)}$ is the number of MSP in the positive (negative) $\alpha(\hat{\Omega},\hat{p},\theta)$ region, and $\mu_{+(-)}$ and $\sigma^2_{+(-)}$ are the mean value and variance of the $\log_{10}\dot{p}/p$ distribution in each region, respectively, 
\begin{eqnarray}
\mu_{+(-)} &=&  \frac{1}{N_{+(-)}}\sum^{N_{+(-)}}_i \log_{10}\left(\frac{\dot{p}}{p}\right)_i, \\
\sigma_{+(-)}^2 &=& \frac{1}{N_{+(-)}}\sum^{N_{+(-)}}_i \left( \log_{10}\left(\frac{\dot{p}}{p}\right)_i - \mu_{+(-)} \right)^2. 
\end{eqnarray}
If we choose correct source position and GW polarization angle, we would obtain nonzero value of $\Delta S$. We obtain a probability distribution of the skewness difference $\Delta S$ for a fixed $\dot{h}$ value from 10,000 realizations. It should be noted that the simulation results are independent of the type of GW sources, if GWs come from a single source. 
Also, because MSPs are distributed isotropically, our result does not depend on the 
source position and GW polarization angle.

\begin{figure}
\centering
\includegraphics[width=60mm, angle=-90]{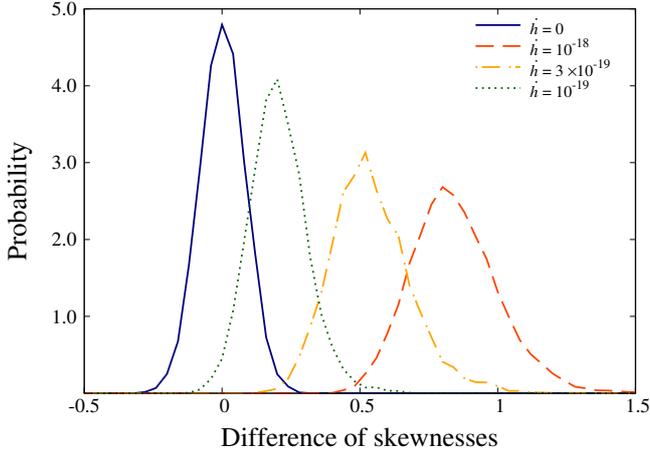}
\caption{Probability distribution of the skewness difference of the $\log_{10}\dot{p}/p$ distributions between the positive and negative $\alpha (\hat{\Omega},\hat{p})$ groups in the case of 3,000 MSPs. The vertical axis represents the probability per unit skewness difference. The navy line is one in the absence of GWs. The red, orange and green lines correspond to the cases with $\dot{h} = 10^{-18}$, $3\times10^{-19}$ and $10^{-19}~\rm{s}^{-1}$ respectively.}
\label{diff-skew3000}
\end{figure}

Let us begin with the results of our fiducial simulations with 3,000 MSPs. This number of MSPs is expected by a future pulsar survey with SKA2 \citep{Keane}. Considering a situation where we have specific targets such as Sgr A${}^*$ and M87, we set the GW source position to the correct values. In addition, the GW polarization angle is also set to the correct value. This is not practical and we discuss this later in this section. Then the bias factor $\alpha(\hat{\Omega},\hat{p},\theta)$ is fixed and mock MSPs are divided into two groups. Fig. \ref{diff-skew3000} shows the probability distribution of the skewness difference. A large $\dot{h}$ value leads to a large skewness difference and the average values are 0.20, 0.54 and 0.84 for $\dot{h} = 10^{-19}$, $3 \times 10^{-19}$ and $10^{-18}$, respectively, while the standard deviations are 0.11, 0.14 and 0.16, respectively. On the other hand, in the absence of GWs, the average value of skewness difference is zero and the standard deviation is $8.2\times 10^{-2}$. A hypothesis of no GWs is rejected for the skewness difference larger than 0.17 at 98\% confidence. Thus, GWs of $\dot{h} \gtrsim 3 \times 10^{-19}$ would be detected with 3,000 MSPs in the absence of other noise contributions.

\begin{figure}
\centering
\includegraphics[width=60mm, angle=-90]{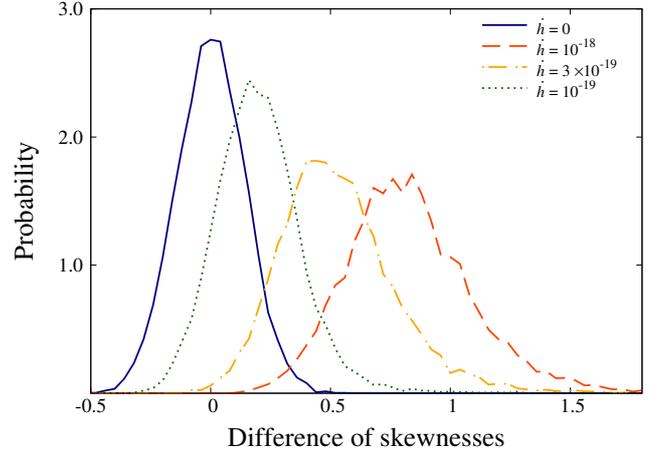}
\caption{Probability distribution in the case of 1,000 MSPs. The line types are the same as Fig. \ref{diff-skew3000}.}
\label{diff-skew1000}
\end{figure}

\begin{figure}
\centering
\includegraphics[width=60mm, angle=-90]{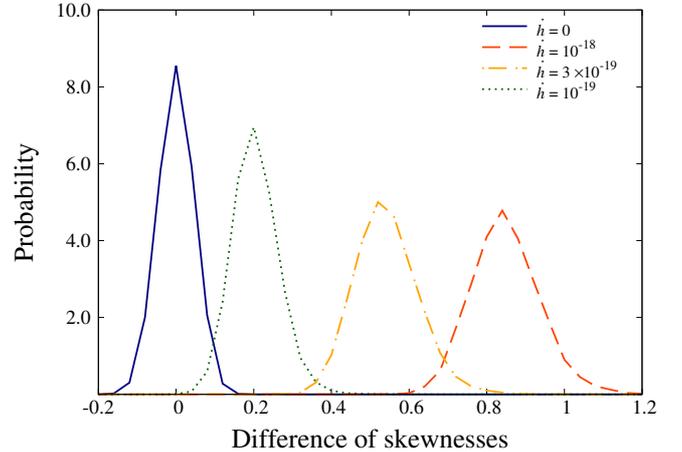}
\caption{Probability distribution in the case of 10,000 MSPs. The line types are the same as Fig. \ref{diff-skew3000}.}
\label{diff-skew10000}
\end{figure}

Figs. \ref{diff-skew1000} and \ref{diff-skew10000} show the probability distributions of the skewness difference for 1,000 and 10,000 MSPs, respectively. The number of MSPs is expected to reach 1,000 by an SKA1 survey. Although 10,000 MSPs are not realistic even with the SKA2, we consider this case in order to investigate the ultimate capability of this method. As the number of MSPs increases (decreases), the standard deviation of the probability distribution decreases (increases) and the sensitivity for GWs is estimated to be about $\dot{h} \sim 10^{-18}$ with 1,000 MSPs and $\dot{h} \sim 10^{-19}$ with 10,000 MSPs. Thus, the sensitivity improves almost proportionally to the number of MSPs.

\subsection{Effect of the intrinsic skewness of the spin-down rate distribution}
In the above analyses, we assumed the intrinsic $\log_{10}\dot{p}_0/p$ distribution is Gaussian. However, as we show below, the $\log_{10}\dot{p}_0/p$ distribution of known MSPs is not Gaussian. To test the Gaussian, we use the Jarque-Bera test. The test statistic $JB$ is defined as
\begin{equation}
JB = \frac{N}{6}\left[ S^2 + \frac{1}{4}(K-3)^2 \right],
\end{equation}
where $K$ is the sample kurtosis and $N$ is the number of samples. If the sample distribution is Gaussian, this test statistic follows the $\chi^2$ distribution with 2 degrees of freedom, and the critical value for $99.5\%$ confidence is 10.6. From the known MSP samples in Fig. \ref{MSP}, we obtain $S = 1.2$ and $K = 5.9$, which leads to $JB = 84.2$ and rejects the Gaussian with more than $99.9 \%$ confidence (for the interpretation of the $\log_{10}\dot{p}_0/p$ distribution, see \citet{Kiziltan}).

\begin{figure}
\centering
\includegraphics[width=60mm, angle=-90]{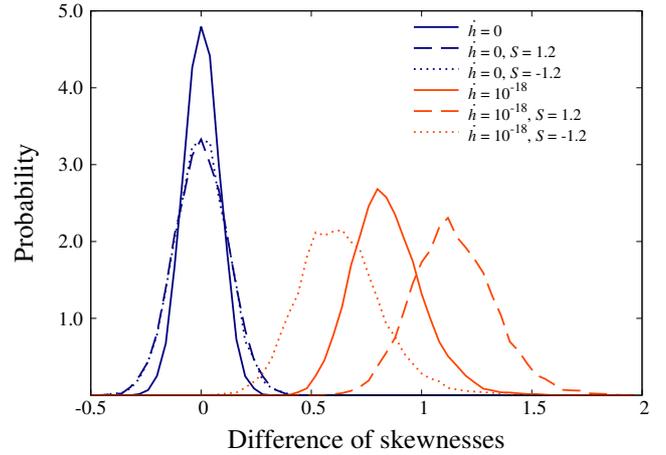}
\caption{Probability distribution of the skewness difference for non-Gaussian intrinsic $\log_{10}\dot{p}_0/p$ distribution with 3,000 MSPs. The navy and red lines are ones with no GWs and $\dot{h} = 10^{-18}~\rm{s}^{-1}$, respectively. The solid, dashed and dotted lines correspond to the cases with intrinsic skewness of 0 (Gaussian), 1.2 and -1.2, respectively.}
\label{dep-skew}
\end{figure}

Thus, we study the effect of non-Gaussian of intrinsic $\log_{10}\dot{p}_0/p$ distribution. To do this, we utilize a generalized normal distribution given by
\begin{equation}
f(x) = \frac{\phi (y)}{\alpha - \kappa (x - \xi)},
\end{equation}
where $\phi (y)$ is the standard normal distribution and $y$ is given by
\begin{eqnarray}
y =
  \left\{
    \begin{array}{ll}
      -\frac{1}{\kappa}\log \left[ 1 - \frac{\kappa (x - \xi)}{\alpha}\right] & (\kappa\ne 0)\\
      \frac{x - \xi}{\alpha} & (\kappa = 0).
    \end{array}
  \right.
\end{eqnarray}
Here, $\xi$, $\alpha$ and $\kappa$ are the location, scale and shape parameters, respectively, and the mean value $\mu$, variance $\sigma^2$ and skewness $S$ are expressed by these parameters.
\begin{eqnarray}
\mu &=& \xi - \frac{\alpha}{\kappa}\left( e^{\kappa^2/2}-1 \right)\\
\sigma^2 &=& \frac{\alpha^2}{\kappa^2}e^{\kappa^2}\left( e^{\kappa^2} - 1 \right) \\
S &=& \frac{3e^{\kappa^2} - e^{3\kappa^2} - 2}{(e^{\kappa^2} - 1)^{3/2}}\rm{sign}(\kappa)
\end{eqnarray}
Assuming that the intrinsic $\log_{10}\dot{p}_0/p$ distribution follows the generalized Gaussian distribution, we perform the same simulations as above with 3,000 MSPs. Fig. \ref{dep-skew} shows the probability distribution of the skewness difference for the intrinsic skewness of -1.2, 0 (Gaussian) and 1.2. In the absence of GWs, the intrinsic skewness increases the standard deviation of the probability distribution of the skewness difference. On the other hand, in the presence of GWs ($\dot{h} = 10^{-18}~\rm{s}^{-1}$), the intrinsic skewness not only widens but also shifts the probability distribution. As can be seen, a positive intrinsic skewness enhances the skewness difference. This is because the median of the intrinsic $\log_{10}\dot{p}_0/p$ distribution is smaller than the average and the number of MSPs with a small value of $\log_{10}\dot{p}_0/p$ increases for a fixed value of the average. Thus a positive intrinsic skewness, which is indicated by the known MSP samples, enhances the detectability of GWs. Contrastingly, a negative intrinsic skewness would reduce the detectability.

\subsection{Precision of the determination of GW polarization angle}
\begin{figure}
\centering
\includegraphics[width=60mm, angle=-90]{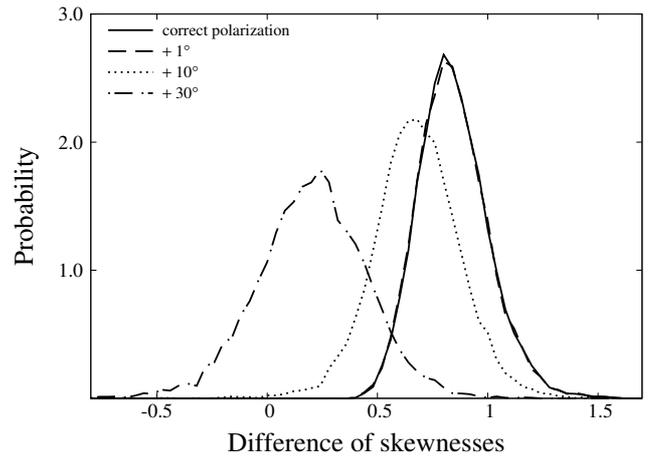}
\caption{Probability distribution of the deviation angle which gives the maximum skewness difference in the case of 3,000 MSPs and $\dot{h} = 10^{-18}~\rm{s}^{-1}$. The solid, dashed and dotted lines correspond to the deviation of $1^\circ$, $10^\circ$ and $30^\circ$, respectively.}
\label{dep-ang}
\end{figure}

\begin{figure}
\centering
\includegraphics[width=60mm, angle=-90]{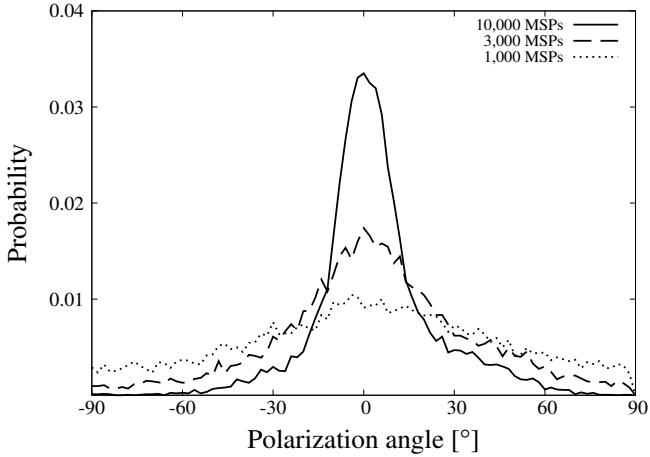}
\caption{Precision of the determination of GW polarization angle in the case of $\dot{h} = 10^{-18}~\rm{s}^{-1}$. The numbers of MSPs are 10,000 (solid), 3,000 (dashed) and 1,000 (dotted).}
\label{res-ang}
\end{figure}

In the simulations shown above, we assumed the correct GW polarization angle is known. If we assume a wrong value of polarization angle, the skewness difference will be reduced. To see this, we perform simulations where the assumed polarization angle is deviated by $1^\circ$, $10^\circ$ and $30^\circ$ from the correct one. Fig. \ref{dep-ang} shows the resultant probability distribution of the skewness difference in the case of 3,000 MSPs and $\dot{h} = 10^{-18}~\rm{s}^{-1}$. From the figure, we see that, if the deviation is more than $10^\circ$, the skewness difference becomes significantly small. In a practical situation, we need to calculate the skewness differences varying the polarization angle while fixing the target position. Then, a polarization angle which gives the maximum skewness difference can be used as an estimate of the correct polarization angle. Fig. \ref{res-ang} shows the probability distribution of the deviation angle which gives the maximum skewness difference for 1,000, 3,000 and 10,000 MSPs and $\dot{h} = 10^{-18}~\rm{s}^{-1}$. The mean is located at $0^\circ$, that is, the estimation is not biased. On the other hand, the standard deviation decreases as the number of MSPs increases: $42^\circ$, $32^\circ$ and $19^\circ$ ($1 \sigma$) for 1,000, 3,000 and 10,000 MSPs, respectively. This can be regarded as the precision of determination of the polarization angle.

\section{Discussion and Summary}\label{discussion}

In this paper, we evaluated the potential of the new detection method for ultra-low frequency GWs ($\lesssim 10^{-10}~{\rm s}^{-1}$) from a single source proposed in \citet{Yonemaru}. This method is based on the statistics of observed spin-down rates of MSPs and GW signal appears as the difference of skewness between the spin-down rate distributions of two MSP groups which are constructed according to the MSP position in the sky. We applied the method to the mock samples of MSPs to estimate the sensitivity. As a result, we found that GWs of $\dot{h} = 3 \times 10^{-19}~\rm{s}^{-1}$ is detectable if we have 3,000 MSPs and the sensitivity is roughly proportional to the number of MSPs.

Let us see more details assuming that the GW source is a SMBH binary. The amplitudes of GWs for a circular binary are given by \citep{Gopakumar}
\begin{equation}
h_A = h_0 \left(\hat{u}_i\hat{u}_j - \hat{r}_i\hat{r}_j \right) e_{A,ij},
\label{amp}
\end{equation}
where, $h_0$ is written by
\begin{equation}
h_0 = \frac{2~(G\mathcal{M})^{5/3}(\pi f_{\rm{GW}})^{2/3}}{c^4R},
\end{equation}
where $G$ is the gravitational constant, $c$ is the speed of light, $\mathcal{M} = (m_1m_2)^{3/5}/(m_1+m_2)^{1/5}$ is the chirp mass of the binary where $m_1$ and $m_2$ are the component masses, $R$ is the distance to the source and $f_{\rm{GW}}$ is the GW frequency. Here, $\hat{r}$ and $\hat{u} = \dot{\hat{r}}$ are the relative position and velocity vectors for the two black holes in the orbit described by the orbital phase $\phi$ which is the angle measured from $\hat{m}$. The time derivatives of the amplitudes is given by
\begin{equation}
\dot{h}_A
= - \frac{1}{2} \dot{h}_0\left(\hat{r}_i\hat{u}_j + \hat{u}_i\hat{r}_j \right) e_{A,ij}
\end{equation}
where $\dot{h}_0 = 2 \pi f_{\rm{GW}} h_0$ and this value could be obtained from the current method. Further, the inclination $\iota$, the orbital phase $\phi$, and $\lambda$ which is the angle between the line of nodes and $\alpha (\hat{\Omega}, \hat{p},\theta) = 0$ are related as,
\begin{equation}
\left( 1 + \cos^2\iota \right)\sin 2\phi\cos 2\lambda - 2\cos\iota\cos 2\phi\sin 2\lambda = 0.
\end{equation}

Let us discuss more details of the pulsar term. As mentioned in section \ref{principle}, we neglected the pulsar term in our simulations. This is because, if the pulsar term for many different pulsars has a random phase, it contributes to the noise and does not induce a bias so that the effect is reduced statistically for a large number of pulsars. Here, we evaluate a frequency range where this assumption is valid. Following \cite{Sesana}, in the case the GW source is a SMBH binary, at the leading order, the GW frequency evolution due to radiation reaction is given by
\begin{equation}
\frac{df_{\rm{GW}}}{dt} = \frac{96}{5}\pi^{8/3} \left( \frac{G\mathcal{M}}{c^3}\right)^{5/3}f_{\rm{GW}}^{11/3}.
\end{equation}
Then, the variation of the GW frequency during the pulse propagation time is
\begin{eqnarray}
\Delta f_{\rm{GW}} &=& \int^t_{t-\tau} \frac{df_{\rm{GW}}}{dt} dt \nonumber \\
&\sim& \frac{df_{\rm{GW}}}{dt} \tau \nonumber \\
&\approx& 6.3\times 10^{-16} \mathcal{M}_{8.5}^{5/3}~f_{100~\rm{yr}}^{11/3}~\tau_{5~\rm{kpc}}~~~\rm{Hz}~,
\end{eqnarray}
where
\begin{eqnarray}
\mathcal{M}_{8.5} &=& \frac{\mathcal{M}}{10^{8.5}~M_\odot} \\
f_{100~\rm{yr}} &=& \frac{f_{\rm{GW}}}{(100~\rm{yrs})^{-1}} \\
\tau_{5~\rm{kpc}} &=& \frac{\tau}{5~{\rm{kpc}}/c~\rm{sec}}~.
\end{eqnarray}
This implies that the GW frequency changes little during the pulse propagation time for $\leq 10^{-10}$ Hz. The Earth and pulsar terms have the same frequency and their phase difference is given by $2\pi f_{\rm{GW}}\tau$. In order for the pulsar term of many different pulsars to have a random phase, the phase difference should be rather large, $2\pi f_{\rm{GW}}\tau \gtrsim \mathcal{O}(0.1)$ for a typical distance $L$. With $L \sim$ 5 kpc in the SKA era and assuming that the angle between the directions to the pulsar and GW source is not so small, the phase difference becomes large for $ \gtrsim 10^{-13}$ Hz. Thus, the pulsar term contributes to a random noise for $10^{-13}-10^{-10}$ Hz. However, the effect of the pulsar term can not be neglected when the directions to the pulsar and GW source coincide, since $\tau = 0$ and the Earth and pulsar terms cancel out. This effect is known as the "surfing effect" \citep{Braginsky} and important when $2\pi f_{\rm{GW}}\tau \ll 1$. This condition can be rewritten, using the angle from the pulsar direction to the GW source $\delta\zeta$,
\begin{equation}
1 - \cos(\delta\zeta) \ll \frac{c}{\pi f_{\rm{GW}}L}.
\end{equation}
For $f_{\rm{GW}} \sim (\rm{a~few}~\times~100~\rm{yrs})^{-1}~\approx~10^{-10}$ Hz and $L = 5$ kpc, the surfing effect is effective for an area of $\sim 127.7~\rm{deg}^2$. Because this is only $\sim 0.3\%$ of the entire sky, this effect is not so significant for this configuration. 
For lower frequency, however, the effect is more significant and a "hole" appears around the GW source in the bias map (Fig. \ref{fig:contour}). Assuming the same typical distance of 5 kpc, this hole covers $\sim$ 10\%, 50\% and 80\% of the sky for GW frequencies of $\sim 3.2\times 10^{-12}$, $7.35\times 10^{-13}$ and $5.0\times 10^{-13}$ Hz, respectively. This would effectively reduce the number of MSPs contributing to the experiment and some reduction of the overall sensitivity.

For lower frequencies $\ll 10^{-13}$ Hz, the pulsar term can no longer be regarded as random noise. In this case, Eq. (\ref{dif metric}) can be written as 
\begin{eqnarray}
\Delta h_A(t,\hat{\Omega}) &=& \dot{h}_A t \left( 1 - e^{2i\pi f_{\rm{GW}}\tau}\right) \nonumber \\
&\simeq& \dot{h}_{Af^2} t \left( 1 + \hat{\Omega}\cdot\hat{p}\right)^2,
\end{eqnarray}
where $\dot{h}_{Af^2} = 2\pi^2 f^2_{\rm{GW}}L^2/c^2~\dot{h}_A$. In the second step, we have used Taylor expansion assuming $2\pi f_{\rm{GW}}\tau \ll 1$. Then the bias factor is given by
\begin{equation}
\alpha(\hat{\Omega},\hat{p}) = \frac{1}{2} \sum_{A = +, \times} \hat{p}^i \hat{p}^j e^A_{ij}\left( 1 + \hat{\Omega}\cdot\hat{p}\right)~\dot{h}_{Af^2}.
\end{equation}
Therefore, we may be able to extract the information of the degenerated value $f^2_{\rm{GW}}\dot{h}_A$ if the distance to the pulsar is measured precisely. Although the effect would be very small as it has additional factor $(f_{\rm{GW}}L/c)^2 \ll 1$, it would be a unique method to probe GWs at such low frequency. 
The quantitative estimation of sensitivity for such GWs should be studied elsewhere in the future.

One should also take into account the observational error in $\log_{10}(\dot{p}/p)$ which propagates to the uncertainty in the skewness difference. The uncertainty of $\Delta S$ can be estimated by considering error propagation and given by
\begin{equation}
\sigma^2_{\Delta S} = \sigma^2_{S_{\alpha +}} + \sigma^2_{S_{\alpha -}},
\end{equation}
where $\sigma_{S_{\alpha +(-)}}$ is uncertainty of the skewness in each region and given by
\begin{eqnarray}
&&\sigma^2_{S_{\alpha +(-)}} =  \left( \frac{3}{N_{+(-)}\sigma_{+(-)}^3} \right)^2 \nonumber \\
&&\times \sum^{N_{+(-)}}_i \left( \left( \log_{10}\left(\frac{\dot{p}}{p}\right)_i - \mu_{+(-)} \right)^2 - \sigma_{+(-)}^2 \right)^2 \sigma^2_{\log_{10}\left(\dot{p}/p\right)_i}\label{error_skew},
\end{eqnarray}
where $\sigma^2_{\log_{10}\left(\dot{p}/p\right)_i}$ is uncertainty in $\log_{10}(\dot{p}/p)$, which 
is typically ${\mathcal{O}}(0.01)$. Taking $\sigma^2_{\log_{10}\left(\dot{p}/p\right)_i} = 0.01$ leads to the uncertainty in the skewness difference of $\sim 5 \times 10^{-2}$. This is much smaller than the value of skewness difference which we are interested here. 

Let us consider some specific targets. A SMBH with mass of $4.0 \times 10^6 M_\odot$ is known to reside in Sgr A${}^*$ and the possibility of the existence of another SMBH has been discussed (e.g. \citet{Oka}). If they are forming a binary and emitting GWs of a frequency $(100~\rm{yrs})^{-1}$, for example, the mass of the second SMBH must be greater than $10^{15} M_\odot$ to be detectable by the current method with 3,000 MSPs. A more interesting target is M87 which has a SMBH with mass of $6.6 \times 10^9 M_\odot$. Constraints on the amplitude of GWs emitted by a milli-pc scale SMBH binary in M87 and other galaxies in the PTA frequency bands has been already studied \citep{Schutz}. While, a possibility of another SMBH outside the AGN has been indicated also for M87 \citep{Batcheldor} and \citet{Yonemaru} discussed GWs from such a potential pc-scale SMBH binary. There, we found that $\dot{h}$ can be as large as $10^{-18}~\rm{s}^{-1}$ when the second SMBH mass is $6.6 \times 10^8 M_\odot$, the eccentricity of the orbit is greater than 0.8 and the black holes are located near the perihelion.

\section*{Acknowledgements}
\addcontentsline{toc}{section}{Acknowledgements}

We thank George Hobbs for useful discussion. KT is supported by Grand-in-Aid from the Ministry of Education, Culture, Sports, and Science and Technology (MEXT) of Japan, No. 15H05896, No. 16H05999, No. 17H01110 and No. 26610048, and Bilateral Joint Research Projects of JSPS.





\begin{thebibliography}{99}


\bibitem[Anholm et al. (2009)]{Anholm}
Anholm, M., Ballmer, S., Creighton, J. D. E., Price, L. R. \& Xavier, S., 2009, Phys. Rev. D, 79, 084030

\bibitem[Batcheldor et al. (2010)]{Batcheldor}
Batcheldor, D., Robinson, A., Axon, D. J., Perlman, E. S. \& Merritt, D., 2010 ApJ, 717, L6

\bibitem[Bertotti et al. (1983)]{Bertotti}
Bertotti, B., Carr B. J. \& Rees M. J., 1983, MNRAS, 243, 945

\bibitem[Braginsky et al. (1990)]{Braginsky}
Braginsky, V. B., Kardashev, N. S., Polnarev, A. G. \& Noviko, I. D., 1990, Nuovo Cimento B, 105, 1141

\bibitem[Champion et al. (2010)]{Champion}
Champion, D. J. et al, 2010, ApJ, 720, L201

\bibitem[Damour \& Taylor (1991)]{Damour}
Damour, T. \& Taylor, J. H., 1991, ApJ, 366, 501

\bibitem[Detweiler (1979)]{Detweiler}
Detweiler, S., ApJ, 234, 1100

\bibitem[Foster \& Backer (1990)]{Foster}
Foster, R. S. \&Backer, D. C., 1990, ApJ, 361, 300

\bibitem[Gopakumar \& Iyer (2002)]{Gopakumar}
Gopakumar, A. \& Iyer, B. R., 2002, Phys. Rev. D, 65, 084011

\bibitem[Jenet et al. (2009)]{Jenet}
Jenet, F. et al, 2009, arXiv:0909.1058

\bibitem[Keane et al. (2015)]{Keane}
Keane, E. F. et al, 2015, arXiv:1501.00056

\bibitem[Kiziltan \& Thorsett (2010)]{Kiziltan}
Kiziltan, B. \& Thorsett, S. E., 2010, ApJ, 715, 335

\bibitem[Kopeikin (1997)]{Kopeikin1997}
Kopeikin, S. M., 1997, Phys. Rev. D, 56, 4455

\bibitem[Kopeikin (1999)]{Kopeikin1999}
Kopeikin, S. M., 1999, MNRAS, 305, 563

\bibitem[Kopeikin \& Potapov (2004)]{Kopeikin2004}
Kopeikin, S. M. \& Potapov, V. A., 1999, MNRAS, 355, 395

\bibitem[Kramer \& Champion (2013)]{Kramer13}
Kramer, M. \& Champion, D. J., 2013, Classical and Quantum Gravity, 30, 4009 

\bibitem[Kramer \& Stapper (2015)]{Kramer15}
Kramer, M. \& Stapper, B., 2015, arXiv:1507.04423

\bibitem[Lodato et al. (2009)]{Lodato}
Lodarto, G., Nayakshin, S., King, A. R. \& Pringle, J. E., 2009, MNRAS, 398, 1392

\bibitem[Manchester et al. (2012)]{Manchester}
Manchester, R. N., et al. 2012, PASA, 30, 17

\bibitem[Milosavljevi´c \& Merritt (2001)]{Milosavljevi´c}
Milosavljevi´c, M. \& Merritt, D., 2001, ApJ, 563, 34

\bibitem[Nice \& Taylor (1995)]{Nice}
Nice, D. J. \& Taylor, J. H., ApJ, 441, 429

\bibitem[Oka et al. (2016)]{Oka}
Oka, T., Mizuno, R., Miura, K. \& Tatekawa, S., 2016, ApJ, 816, L7

\bibitem[Phinney (1993)]{Phinney}
Phinney, E. S., 1993, ASPC, 50, 141

\bibitem[Potapov et al. (2003)]{Potapov}
Potapov, V. A., Ilyasov, Yu. P., Oreshko, V. V. \& Rodin, A. E., 2003,  Astron. Lett., 29, 241

\bibitem[Rong et al. (1999)]{Rong}
Rong, J., Xiao, N. \& Tan, L., 1999, Science in China Series, A-Math, 42, 444

\bibitem[Schutz \& Ma (2016)]{Schutz}
Schutz, K. \& Ma, C., 2016, MNRAS, 459, 1737

\bibitem[Sesana \& Vecchio (2010)]{Sesana}
Sesana, A., \& Vecchio, A., 2010, Phys. Rev. D, 81, 104008

\bibitem[Shannon et al. (2015)]{Shannon}
Shannon, R. M., et al, 2015, Science, 349, 1522

\bibitem[Shklovskii (1970)]{Shklovskii}
Shklovskii, I. S., 1970, Soviet Astronomy, 13, 562


\bibitem[Yonemaru et al. (2016)]{Yonemaru}
Yonemaru, N., Kumamoto, H., Kuroyanagi, S., Takahashi, K. \& Silk, J., 2016, PASJ, 68, 106

\end{thebibliography}






\bsp	
\label{lastpage}
\end{document}